\documentclass{article}

\usepackage[english]{babel}

\usepackage[letterpaper,top=2cm,bottom=2cm,left=3cm,right=3cm,marginparwidth=1.75cm]{geometry}

\usepackage{amsmath}
\usepackage{graphicx}
\usepackage[colorlinks=true, allcolors=blue]{hyperref}

\newcommand{\apj}{Astrophys. J.}

\newcommand{\mnras}{Mon. Not. R. Astron. Soc.}
\newcommand{\nat}{Nature}

\newcommand{\araa}{Ann. Rev. of Astron. Astrophys.}

\newcommand{\aap}{Astron. Astrophys.}

\title{Planes of satellites, no longer in tension with $\Lambda$CDM?}
\author{Laura V. Sales$^{a}$$^{*}$ and Julio F. Navarro$^{b}$ \\
        \small $^{a}$Department of Physics and Astronomy, University of California, Riverside, CA, 92521, USA\\
        \small $^{b}$Department of Physics and Astronomy, University of Victoria, Victoria, BC V8P 5C2, Canada\\
        \small $^{*}$ \tt{lsales@ucr.edu}
}

\begin{document}
\maketitle



\begin{center}
{\bf The arrangement of dwarf galaxies in a thin plane surrounding the Milky Way has been thought to contradict the prevailing cosmological model of cold dark matter in the Universe. New work suggests that this arrangement may just be a temporary alignment, bringing our galaxy back into agreement with theoretical expectations once the radial distribution of satellites is taken into account.}
\end{center}

The $\Lambda$CDM cosmogony predicts that the spatial distribution of satellites around galaxies like the Milky Way should be modestly anisotropic, an evolving configuration set  by the continuous accretion of satellites from the cosmic web and where angular momentum plays a negligible role. By contrast, the eleven most luminous satellites of the Milky Way outline a thin plane whose minor axis coincides with the orbital angular momentum of many of them. It has been argued that such planes are highly unlikely in $\Lambda$CDM, setting up a tension that has remained so far unresolved. The tension is unwarranted: the thinness of the Milky Way plane is transitory and boosted by the chance alignment of its two most distant satellites, and the probability of finding such configurations in $\Lambda$CDM simulations has been severely underestimated in earlier work because of improper accounting of the radial distribution of simulated satellites.

Any individual system scrutinized in detail is bound to reveal ``oddities", or unexpectedly peculiar traits. In some cases these are evidence for physical effects beyond our current understanding; in others, they are simply coincidences of little import. The ``plane of satellites" tension in the Milky Way is one example of this dilemma \cite{Bullock2017,Pawlowski2021,Sales2022}. Because of our location, we have been able to study the position and motions of satellites in the Milky Way in great detail: the satellite plane is not only much thinner than one would naively expect, but also coincides with the orbital plane of many of the satellites that define it. Although the presence of rotating satellite planes around other galaxies, like M31 (Andromeda) \cite{Ibata2013} and NGC 5128 (Centaurus A) \cite{Muller2021}, has also been argued for, the evidence in those cases is much weaker and inconclusive due mainly to  lack of detailed kinematic data (only line-of-sight velocities are available for external galaxies).

Thin, co-rotating planar structures are not uncommon in astrophysics, but thoroughly unexpected when it comes to galactic satellites. Astrophysical disks form as a result of energy dissipation and angular momentum conservation, a process that takes many orbital times to complete. None of these conditions apply to the Milky Way satellite plane: there is no known mechanism for satellites to dissipate their orbital energy, and many Milky Way satellites have completed at most a few orbits around their host. A rotating thin plane of satellites is therefore completely unexpected, not only in $\Lambda$CDM, but also in most, if not all, potentially viable models of galaxy formation \cite{BK2021}. 

Prior work has used $\Lambda$CDM simulations to quantify the probability of finding a plane of satellites like that of the Milky Way, and concluded that the chances are vanishingly small when comparing to the distribution of subhalos \cite{Kang2005,Pawlowski2018}. Does this mean that we should  overhaul our current understanding and search for physics ``beyond the standard model" of galaxy formation? Or should we instead critically re-examine the likelihood of such configurations arising in $\Lambda$CDM? 

Sawala et al \cite{Sawala2022} argue forcefully for the latter option. Their analysis demonstrates that the thinness of the Milky Way satellite plane is driven mainly by the two most distant objects (Leo I and Leo II), a consequence of the radial distribution of the small sample of satellites (11 in total) considered. In addition, the plane's thickness is transient \cite{Helmi2018}, as Leo I and Leo II are actually on different orbits aligned by chance at the present time:  the configuration of the same satellites, considered one Gyr before or after today would have been much closer to the expectations from $\Lambda$CDM simulations. 

Speaking of expectations, Sawala et al also argue that the probability of finding planes of satellites like that in the Milky Way has been severely underestimated in earlier work. Their simulation analysis attempts to correct for the artificial disruption of satellites due to limited numerical resolution \cite{vandenBosch2018,Errani2021}, and to analyze systems where, after such corrections, the radial distribution of satellites matches approximately that of the Milky Way. They report that the likelihood of having a satellite system as planar as that of the Milky Way is about $50$ times larger than previously thought, reaching $\sim 2\%$ of all simulated Milky-Way like objects, a conclusion also reached by a recent independent study \cite{Pham2022}. Our Galactic plane of satellites might not be extremely common, but it seems eminently possible.

Although  the plane of satellites in the Milky Way might not be a fundamental challenge to $\Lambda$CDM, that does not mean that there is nothing to be learned from studying it in detail. What is special about the assembly history and local environment of the Galaxy that led to such peculiar configuration? If it is a result of a preferential accretion direction steered by  filaments of the cosmic web \cite{Libeskind2005}, there may very well be signatures of such filaments outside the virial boundary of the Milky Way halo. There is little evidence of such structure in presently available data, but there are also strong reasons to believe that our inventory of dwarf galaxies in the Local Group is woefully incomplete \cite{Tollerud2008,Fattahi2020}.

The advent of facilities such as the Rubin Observatory and the Roman Space Telescope gives reasons to be optimistic about substantial progress in the near future. If simulation predictions are correct, such facilities should turn into dwarf galaxy discovery machines that would enable us to trace and characterize structures throughout our Local Group. It is important that such progress is also accompanied with more precise predictions from simulations about the distribution of satellites inside and outside the Galactic virial radius. It is not enough for simulators to argue that ``this happens" in $\Lambda$CDM; we would like to know also ``why" and ``how" it happens. After all, if our Galaxy is an outlier among similar systems it would be good to understand the origin of its peculiarities and to make concrete predictions that would allow the invoked explanations  to be falsified.  This is especially important because, as we dig deeper into our local patch of the Universe, the planes of satellites problem is unlikely to be the only riddle that will need puzzling out. 

\begin{figure}[ht]
\centering
\hspace{-1 cm}
\includegraphics[width = 0.86 \linewidth]{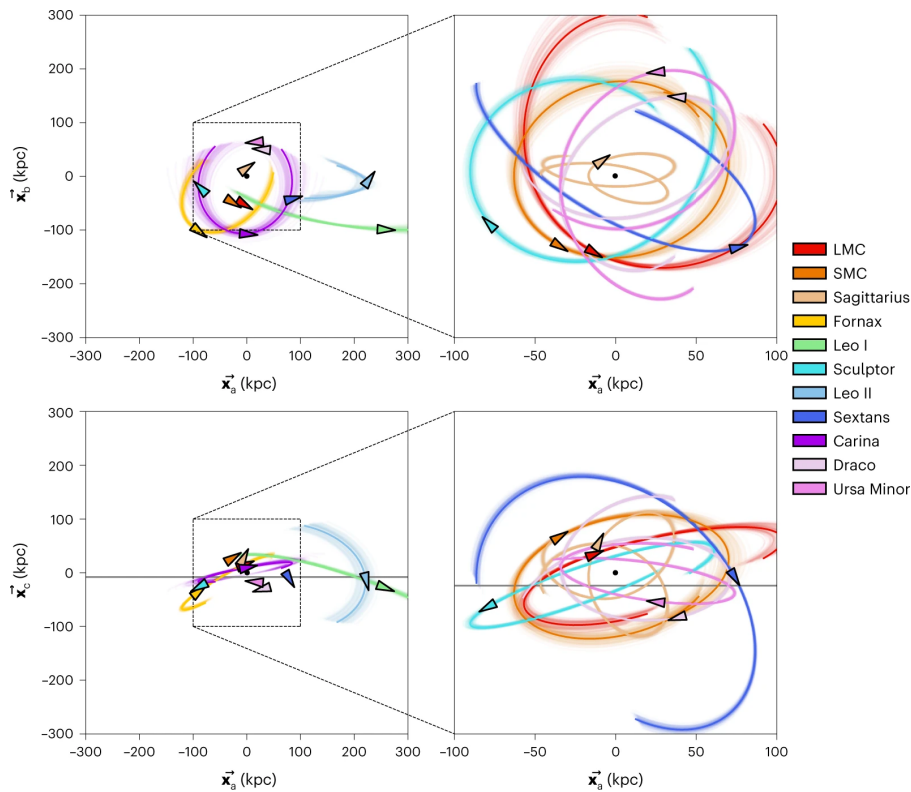}
\caption{
\textbf{The unusual thinness of the plane of satellites in the Milky Way is heavily defined by two of the most distant satellites, Leo I and Leo II}, adapted from ref. \cite{Sawala2022}.
Upper and lower rows show the face-on and edge-on projections of the satellite plane, respectively, as calculated with the inertia tensor of the $11$ Milky Way satellites in the legend.  The radially concentrated distribution of the inner satellites contrasts with the presence of only $2$ relatively external dwarfs, Leo I and Leo II with distances larger than 100 kpc, mostly responsible for the very extreme axis ratios calculated for the satellite plane. Moreover, as shown by the colored arcs depicting the orbit evolution expected during 1 Gyr, this extreme alignment is only temporary, becoming a thicker distribution quickly as LeoII moves perpendicular to the edge-on plane. 
}
\end{figure}

\section*{Competing interests}

The authors declare no competing interests.


\end{document}